\documentclass[%
preprint,
showpacs,
amsmath,amssymb,
aps,
]{revtex4-1}

\usepackage[dvipdfmx]{graphicx,color}
\usepackage{bm}
\usepackage{setspace}

\begin{document}

\preprint{}

\title{P\'{o}lya urn with memory kernel and \\  
asymptotic behaviors of autocorrelation function}

\author{Shintaro Mori}
\email{shintaro.mori@hirosaki-u.ac.jp}
\affiliation{
Department of Mathematics and Physics,
Faculty of Science and Technology, 
Hirosaki University, \\ 
Bunkyo-cho 3, Hirosaki, Aomori 036-8561, Japan
}

\author{Masato Hisakado}
\email{hisakadom@yahoo.co.jp}
\affiliation{
  Nomura Holdings Inc., \\ 
  Otemachi 2-2-2, Chiyoda-ku, Tokyo 100-8130, Japan
}%

\author{Kazuaki Nakayama}
\email{nakayama@math.shinshu-u.ac.jp}
\affiliation{
Department of Mathematical Sciences,
Faculty of Science, Shinshu University, \\
Asahi 3-1-1, Matsumoto, Nagano 390-8621, Japan
}%

\date{\today}

\begin{abstract}
  P\'{o}lya urn is a stochastic process in which balls are randomly drawn
  from an urn of red and blue balls, and balls of the same color as the drawn
  balls are added. The probability of a ball of a certain color being drawn
  is equal to the percentage of balls of that color in the urn.
  We introduce arbitrary memory kernels to modify this probability.
  If the memory kernel decays exponentially, it is a stationary process and is mean-reverting.
  If the memory kernel decays by a power-law, a phase transition occurs and the asymptotic
  behavior of the autocorrelation function changes.
  An auxiliary field variable is introduced to transform the process Markovian and the field obeys 
  a multivariate Ornstein-Uhlenbeck process. 
  The exponents of the power law are estimated for the decay of the leading and subleading terms of
  the autocorrelation function. It is  shown that the power law exponents changes discontinuously
  at the critical point.
\end{abstract}

\maketitle


\section{\label{sec:intro}Introduction}
Econophysics and socio-physics are fascinating research fields
in statistical physics\cite{Mantegna:2008,Galam:2008,Castellano:2009}.
In particular, herding or the tendency to follow the majority
has attracted many researchers as it
plays a crucial role in the understanding of
the phenomena\cite{Arthur:1989,Bikchandani:1992,Cont:2000,Mori:2016}.
Many types of probabilistic models
have been proposed to model herding and  
one example is the ant recruitment model, which
describes the intermittent oscillation of the ants
faced with two identical food sources\cite{Kirman:1993}.
It incorporates a simple herding mechanism where a
randomly chosen ant chooses one of the two food sources with
a probability that is proportional to the 
numbers of ants who have chosen the food sources.

A variant of the ant recruitment model has been adopted as
a model for the intermittent switching phenomena of trading strategy,
trend-follower and contrarian, in the modeling of the time-series data
of stock and foreign exchange markets\cite{Lux:1995}.
Another application is the 
collective bankruptcy of assets\cite{Hisakado:2020}.
In the modeling of the time-series data of defaults
of assets, the estimation of the default probability (PD)
and the default correlation 
is crucial as it affects the risk estimation of the investment.
In addition to the default correlation for the defaults in
the same year, it is necessary to model the correlation of
the defaults in different years. The beta-binomial distribution model
with memory kernel is a model for both correlations and it is
tractable compared to the established model based on the correlation of
asset values\cite{Pluto:2011,Hisakado:2021}.

The fascinating point of the beta-binomial distribution model with memory
kernel is its phase transition, and its essence is grasped by a
variant of the P\'{o}lya urn\cite{Polya:1931},
P\'{o}lya urn with memory kernel\cite{Hisakado:2020}.
If the memory kernel decays with the power law, the power-law index $\gamma$
determines the phases of the model.
If $\gamma>1$, the autocorrelation function (ACF)
shows power-law decay with exponent $\delta=\gamma$.
For $\gamma<1$, the process becomes nonstationary and 
$\delta=0$. 
If $\gamma$ approaches 1, $\delta$ changes discontinuously
to a value $\delta_c$.
The order parameter of the phase transition is the limit value of the ACF.
In the nonstationary phase with $c>0$ for $\gamma<1$,
the memory of the history remains forever and the estimation
of the PD and other parameters
becomes impossible even if the length of the time series is infinite.

This study addresses the phase transition of the P\'{o}lya
urn with an arbitrary memory
kernel using the stochastic differential equation (SDE).
The original P\'{o}lya urn is a Markov process as the probability
that the color of a newly added ball is red
depends only on the present number of red balls in the urn.
The P\'{o}lya urn with exponential decay memory kernel can also be
described as a Markov process when one focuses on the
probability that the color of a newly added ball is red.
However, if the memory kernel
is a power-law decay one, it is necessary to introduce an infinite number of
auxiliary variables to make the process Markov. The
auxiliary field obeys the SDE with a common Wiener process.
Using the formulation, we derive the self-consistent equation
for the ACF. By studying the equation, we succeed to derive
the power-law exponents of
the leading and subleading terms of the ACF.

Our analysis is very similar to the one that studies the
Hawkes process using a stochastic differential
equation\cite{Kanazawa:2020,Kanazawa:2020-2}.
The Hawkes process is a non-Markov self-excited
point process\cite{Hawkes:1971}.
The Hawkes process
shows a stationary--nonstationary
phase transition when the branching ratio exceeds one. 
The critical and near-critical behaviors of
the probability density function of the intensity have
been studied by solving the Master equation\cite{Kanazawa:2020}.

In our study of the P\'{o}lya urn with memory kernel,
we focus on the phase transition.
Two universality classes are known for the phase transitions
of non-linear P\'{o}lya urns\cite{Mori:2015-2,Nakayama:2021},
which have the same  scaling structure of order parameter 
as that of the absorption state phase transitions\cite{Hinrichsen:2000}.
Our goal is to  understand the nature of the phase transition
of the "linear" P\'{o}lya  urn with power-law memory kernel
 based on the behaviors of the order parameter. 
The organization of the paper is as follows.
Section \ref{sec:model} defines the model.
We review some results of the P\'{o}lya urn.
In section \ref{sec:method},
we study the model using the SDE.
We start from the exponential decay memory kernel model.
The ACF decays exponentially and  the process is stationary.
Next we study the power-law decay  memory kernel model.
We introduce an auxiliary field to make the process Markovian and
we derive the self-consistent equation for the ACF.
Section \ref{sec:results} shows the results about the asymptotic behavior
of the ACF for the power-law decay memory kernel model.
We study the power-law exponents
of the leading and sub-leading terms of the ACF.
In section \ref{sec:conclusion}, we summarize the results.

\section{\label{sec:model}Model}

The P\'{o}lya urn is a binary stochastic process
$X(t)\in \{0,1\},t=1,2,\cdots$ and the probability that $X(t+1)$ takes
the value 1 is 
\begin{equation}
P(X(t+1)=1|X(s),s=1,\cdots,t)
=\frac{\alpha+\sum_{s=1}^{t}X(s)}{\alpha+\beta+\sum_{s=1}^{t}1}. \label{eq:polya}
\end{equation}
$\alpha,\beta>0$ are parameters and can be interpreted as the
initial numbers of red (1) and blue (0) balls in the urn\cite{Polya:1931}.
They are parameters and are not necessarily restricted to be integers.
We denote $\alpha+\beta$ as $\theta$.

We define the P\'{o}lya urn with an arbitrary memory kernel
$d(t),t=0,1,\cdots$\cite{Hisakado:2020}.
We denote the cumulative sum of $d(t)$ as $D_{d}(t)$
\[
D_{d}(t)=\sum_{s=1}^{t}d(s-1),D_{d}(0)=0
\]
and the weighted sum of $X(t)$ with weights $d(t)$ as $S_{d}(t)$,
\[
S_{d}(t)=\sum_{s=1}^{t}X(s)d(t-s)\,,\,S_{d}(0)=0.
\]
$Z_{d}(t)$ is then defined as
\[
Z_{d}(t)=\frac{S_{d}(t)}{D_{d}(t)}\,,\,Z_{d}(0)=0,Z_{d}(1)=X(1).  
\]
The probability that $X(t+1)$ takes the value 1 is
\begin{equation}
P(X(t+1)=1|X(s),s=1,\cdots,t)=\frac{\alpha+S_{d}(t)}{\theta+D_{d}(t)}
=\frac{\alpha+D_{d}(t)Z_{d}(t)}{\theta+D_{d}(t)}.\label{eq:model}
\end{equation}
If one chooses $d(t)=1,t=0,1,\cdots$, $D_{d}(t)=t$ and
$S_{d}(t)=\sum_{s=1}^{t}X(s)$, so that (\ref{eq:model}) reduces
to (\ref{eq:polya}). 
The probabilistic rule of (\ref{eq:model}) is linear
in $X(s)$, the expected value of $X(t)$ is 
$E(X(t))=\alpha/\theta$.

We are interested in the response of $X(t+1)$
by the change in the initial value $X(1)$.
The ACF $C(t)$ is defined as 
the covariance of $X(1)$ and $X(t+1)$ divided by the variance of
$X(1)$ as,
\[
C(t)\equiv \frac{E(X(1)X(t+1))-(\alpha/\theta)^2}{V(X(1)}.
\]
We have another expression for $C(t)$ as \cite{Mori:2015-2},
\[
C(t)=E(X(t+1)|X(1)=1)-E(X(t+1)|X(1)=0).
\]
$X(2)$ depends on $X(1)$ and 
$E(X(2)|X(1))=(\alpha+d(0)X(1))/(\theta+d(0))$, we have
\[
C(1)=\frac{d(0)}{\theta+d(0)}.
\]
The order parameter $c$ of the phase transition of the stochastic process
is defined as the limit value of $C(t)$, $c=\lim_{t\to\infty}C(t)$.
The conditional expected value of $X(t+1)$ under the condition for $X(1)$
is the average probability that $X(t+1)$ takes the value 1 under the same condition. 
\[
E(X(t+1)|X(1))=E(P(X(t+1)=1)|X(1))
=\frac{\alpha+\sum_{s=1}^{t}E(X(s)|X(1))d(t-s)}{\theta+D_{d}(t)}.
\]
As $C(t)=E(X(t+1)|X(1)=1)-E(X(t+1)|X(1)=0)$, we have
the recursive equation for $C(t)$ as,
\begin{equation}
C(t)=\frac{\sum_{s=1}^{t}C(s-1)d(t-s)}{\theta+D_{d}(t)}
=\frac{d(t-1)+\sum_{s=2}^{t}C(s-1)d(t-s)}{\theta+D_{d}(t)}.\label{eq:self}  
\end{equation}
One can compute $C(t),t\ge 2$ using this equation 
and the initial condition $C(1)=d(0)/(\theta+d(0))$.

There are  several choices for the memory kernel $d(t),t=0,\cdots$.
\begin{enumerate}

\item P\'{o}lya urn, $d(t)=1,t\ge 0$.

As we have noted before, if one choose $d(t)=1,t=0,\cdots$, 
the model is the P\'{o}lya urn. 
We denote $Z_{d}(t)$ for this case as $Z_{\infty}(t)$.
$tZ_{\infty}(t)$ obeys a binomial distribution with
shape parameters $\alpha,\beta$\cite{Polya:1931}.  
The expected value of $X(t)$ is $E(X(t))=\alpha/\theta$.
$Z_{\infty}(\infty)$ obeys the beta distribution 
and  the variance is
$\frac{1}{\theta+1}\cdot \frac{\alpha\beta}{\theta^2}$.
When one is interested in the dynamics of
$X(t)$ for the initial condition $X(1)=x$,
it is only necessary to
alter the parameters $\alpha,\beta$ and $\theta$
as $\alpha+x,\beta+(1-x)$ and $\theta+1$, respectively.
The conditional expected value $E(X(t+1)|X(1)=x)$
is $(\alpha+x)/(\theta+1)$ for $t\ge 1$ and ACF $C(t)$
becomes constant.
\begin{equation}
C(t)=\frac{1}{\theta+1},t\ge 1  \label{eq:C_polya}.
\end{equation}
The process is not stationary and not mean-reverting.
The conditional variance of $Z_{\infty}(\infty)$ for
$X(1)=x$ is
\begin{equation}
V(Z_{\infty}(\infty)|X(1)=x)=
\frac{1}{\theta+2}\cdot\frac{(\alpha+x)(\beta+(1-x))}{(\theta+1)^2}
\label{eq:V_polya}.
\end{equation}

\item Finite memory kernel, $d(t)=1,t<r$ and $d(t)=0,t\ge r$.
  
There is a cutoff $r$ in the memory kernel, and $d(t)=1$ for $0\le t<r$,
and $d(t)=0$ for $t\ge r$. For $t>r$, $X(t)$ is influenced by the most recent
$r$ variables $X(s),s=t-1,\cdots,t-r$. For $t\le r$, 
$X(t)$ are influenced from all past variables $X(s),s=1,\cdots,t-1$. 
$Z_{d}(t)$ is defined as
\[
Z_{d}(t)
=\left\{
\begin{array}{cc}
\frac{1}{r}\sum_{s=t-r+1}^{t}X(s) & t\ge r, \\
\frac{1}{t}\sum_{s=1}^{t}X(s) & t<r.
\end{array}
\right.
\]
The probability that $Z_{d}(t),t>r$ changes by $\pm 1/r$ for the condition $Z_{d}(t)=z_{d}$ is given as
\begin{eqnarray}
P(\Delta Z_{d}(t)=+1/r|Z_{d}(t)=z_{d})&=& P(X(t-r+1)=0|Z_{d}(t)=z_{d})\cdot P(X(t+1)=1|Z_{d}(t)=z_{d}),
  \nonumber \\
P(\Delta Z_{d}(t)=-1/r|Z_{d}(t)=z_{d})&=& P(X(t-r+1)=1|Z_{d}(t)=z_{d})\cdot P(X(t+1)=0|Z_{d}(t)=z_{d}).
\nonumber  
\end{eqnarray}
If one adopts the mean field approximation as
$P(X(t-r+1)=1|Z_{d}(t)=z_{d})=z_d(t)$ and $P(X(t-r+1)=0|Z_{d}(t)=z_{d})=1-z_d(t)$, the probabilities
can be written as
\begin{eqnarray}
P(\Delta Z_{d}(t)=+1/r|Z_{d}(t)=z_{d})&=&(1-z_{d}) \cdot \frac{\alpha+rz_{d}(t)}{\theta+r}  \nonumber \\
P(\Delta Z_{d}(t)=-1/r|Z_{d}(t)=z_{d})&=&z_{d} \cdot \frac{\beta+r(1-z_{d}(t))}{\theta+r}
\nonumber  
\end{eqnarray}
By the approximation, the model becomes the Kirman's ant
colony model\cite{Kirman:1993}.
In the Kirman's ant colony model,
there are two food sources: $0,1$ and $r$ ants.
The numbers of ants that have chosen food sources 1 and 0 are $n_{1}$ and
$n_{0}$, respectively.
A randomly selected ant chooses food sources 1 and 0 with
a probability that is linear in $n_{1}$ and $n_{0}$ among $r-1$ ants.
The probability that the ratio $z=n_1/r$ changes $\pm 1/r$ is given as
\begin{eqnarray}
P(\Delta z=+1|n_{1}(t)/r=z)&=& (1-z)\cdot \frac{\alpha+(r-1)z}{\theta+(r-1)},\nonumber \\
P(\Delta z=-1/r|n_{1}(t)/r=z)&=& z\cdot \frac{\beta+(r-1)(1-z)}{\theta+(r-1)}. \nonumber
\end{eqnarray}
The prefactors in the right hand side of the equations are
the probabilities that
the randomly selected ants are from food sources 0 and 1, respectively.
The remaining terms are the probabilities that the selected ant
chooses food sources 1 and 0, respectively. 
The $\alpha,\beta$ are noises in the decision of the ants.
The P\'{o}lya urn with finite memory kernel 
corresponds to  the ant colony model in the mean filed approximation.

The stationary distribution of 
$n_1$ is a beta binomial distribution,
with shape parameters $\alpha,\beta$.
$C(t)$ decays exponentially as $C(t)\sim \exp(-t/\xi)$, which defines
the correlation length $\xi$.
$\xi$  depends on the length of
the memory $r$ and diverges in the limit $r\to\infty$\cite{Mori:2015}.

\item Exponential decay memory kernel, $d(t)=e^{-rt}$.

A previous study reveals that $C(t)$ decays exponentially as
$C(t)\sim e^{-t/\xi}$ for $t>>1$\cite{Hisakado:2020}.
The correlation length $\xi$ is determined by $r$,
\[
1/\xi=\ln \left(\frac{\theta(1-e^{-r})+1}{\theta e^{-r}(1-e^{-r})+1}\right)>0,
\]
and $\xi$ diverges in the limit $r\to 0$.
In the limit $r\to \infty$, we have $d(t)=0$ for $t>0$.
$X(t+1)$ depends only on $X(t)$, which corresponds to the
correlated random walk\cite{Bohm:2000}.
  
\item Power-law decay memory kernel, $d(t)=(1+t)^{-\gamma}$.

The model shows a phase transition at $\gamma=1$\cite{Hisakado:2020}.
By the numerical study of the ACF $C(t)$,  it was conjectured that 
the asymptotic behavior of $C(t)$ is  
\[
C(t)\sim\left\{
  \begin{array}{cc}
    ct^{-\gamma} & \gamma>1, \\
    ct^{-\delta_c} & \gamma=1, \\
    ct^0    & \gamma<1.
  \end{array}  
  \right.
\] 
The critical power-law exponent $\delta_c$ at $\gamma=1$ is given as
the solution of the next equation,
\begin{equation}
\theta =\lim_{t\to\infty}\left( \int_{1/(t+1)}^{t/(t+1)}x^{-\delta_c}(1-x)^{-1}dx-\ln t\right).
\label{eq:critical}
\end{equation}

\end{enumerate}

\section{\label{sec:method}Stochastic differential equation}

We study the P\'{o}lya urn with memory
kernel using the method of stochastic approximation
\cite{Gardiner:2009,Renlund:2010}.
We start from the exponential decay memory kernel model.
For the reader's convenience, we summarize the results
of the analysis of  the original P\'{o}lya urn using the SDE in Appendix A.

\subsection{Exponential decay memory kernel model : $d(t)=e^{-rt}$}

We denote $D_{d}(t),S_{d}(t),Z_{d}(t)$ 
for $d(t)=e^{-rt}$ as
$D(r,t)=\sum_{s=1}^{t}e^{-r(t-s)}$,$S(r,t)=\sum_{s=1}^{t}X(s)e^{-r(t-s)}$, 
$Z(r,t)=S(r,t)/D(r,t)$, respectively.
We decompose $S(r,t+1)$ and $D(r,t+1)$ as 
\begin{eqnarray}
  D(r,t+1)&=&D(r,t)e^{-r}+1,\nonumber \\
  S(r,t+1)&=&\sum_{s=1}^{t+1}X(s)e^{-r(t+1-s)}=S(r,t)e^{-r}+X(t+1).  \nonumber
\end{eqnarray}
$Z(r,t+1)$ can be decomposed as
\begin{equation}
Z(r,t+1)=\frac{S(r,t+1)}{D(r,t+1)}
=Z(r,t)+\frac{X(t+1)-Z(r,t)}{D(r,t+1)} \label{eq:Z_r}.
\end{equation}
For $t>>1$, $D(r,t)=(1-e^{-rt})/(1-e^{-r})\simeq 1/(1-e^{-r})$
and we write $D_{r}=1/(1-e^{-r})$.
The conditional expected value and the variance of
$Z(r,t+1)$ for $Z(r,t)=z_r$ is
\begin{eqnarray}
  E(Z(r,t+1)|Z(r,t)=z_r)-z_r&=&\frac{E(X(t+1)|Z(r,t)=z_r)-z_r}{D(r,t+1)}\simeq \frac{\alpha-\theta z_r}{D_r(D_r+\theta)}
  \nonumber \\
  V(Z(r,t+1)|Z(r,t)=z_r)&=& \frac{1}{D(r,t+1)^2}V(X(t+1)|Z(r,t)=z_r)
  \nonumber \\
  &\simeq& \frac{1}{D_{r}^2}\cdot \frac{(\alpha +D_{r}z_{r})(\beta+D_{r}(1-z_r))}
         {(D_{r}+\theta)^2}. \nonumber 
\end{eqnarray}
Based on the assumption
that ACF decays with $t$ rapidly, $Z(r,t)$ fluctuates around
its mean value and $E(Z(r,t))=\alpha/\theta$, 
we approximate the conditional variance of
$Z(r,t+1)$ for $Z(r,t)=z_r$ by replacing $z_{r}$ with $\alpha/\theta$
as,
\[
V(Z(r,t+1)|Z(r,t)=z_r)\simeq \frac{B^2}{D_{r}^2},B^2=\frac{\alpha\beta}{\theta^2}.
\]
One reads the drift and diffusion terms from the conditional
expected value and the variance, and the SDE becomes
\[
dZ(r,t)=A_r(\alpha-\theta Z(r,t))dt+B_rdW_t
\]
$A_{r},B_{r}$ are defined  as,
\[
A_r=\frac{1}{D_r(D_r+\theta)},B_r=\frac{B}{D_r}.
\]
It is the Ornstein-Uhlenbeck process.
The condition of mean-reverting $A_r \theta >0$ is satisfied 
and $Z(r,t)$ fluctuates around $\alpha/\theta$\cite{Gardiner:2009,Vatiutipong:2019}.

The solution for the SDE with the initial condition $Z(r,1)=X(1)=x$ is
\[
Z(r,t)=\frac{\alpha}{\theta}(1-e^{-\theta A_r(t-1)})+x\cdot e^{-\theta A_r(t-1)}
  +B_r e^{-\theta A_r(t-1)}\int_{1}^{t}e^{\theta A_r(s-1)}dW_s. 
\]
The conditional expected value and the variance of $Z(r,t)$
for the initial condition $X(1)=x$ are
\begin{eqnarray}
  E(Z(r,t)|X(1)=x)&=&\frac{\alpha}{\theta}(1-e^{-\theta A_r(t-1)})+x\cdot e^{-\theta A_r(t-1)}, \nonumber \\
  V(Z(r,t)|X(1)=x)&=& \frac{B_r^2}{2\theta A_r}(1-e^{-\theta A_{r}(t-1)})
  \simeq \frac{B_r^2}{2\theta A_r} \nonumber .
\end{eqnarray}
The conditional variance  does not depend on the initial condition.
For $r<<1$, $D_{r}\simeq 1/r$, $A_r\simeq 1/D_r^2=r^2, B_{r}^2\simeq r^2 B^2$
and $V(Z(r,t)|X(1)=x)\simeq B^2/(2\theta)$.
The ACF is estimated as
\begin{eqnarray}
C(t)&=&E(X(t+1)|X(1)=1)-E(X(t+1)|X(1)=0) \nonumber \\
&=&E(P(X(t+1)=1)|X(1)=1)-E(P(X(t+1)=1)|X(1)=0) \nonumber \\
&=&\frac{D_{r}E(Z(r,t)|Z(1)=1)+\alpha}{D_{r}+\theta}
-\frac{D_{r}E(Z(r,t)|X(1)=0)+\alpha}{D_{r}+\theta} \nonumber \\
&=&\frac{D_r}{D_r+\theta}e^{-\theta A_r(t-1)}=\frac{D_r}{D_r+\theta}e^{-\frac{\theta}{(D_{r}+\theta)D_{r}}t}. \nonumber  
\end{eqnarray}
$C(t)$ decays exponentially and
$C(t)\propto e^{-\theta r^2 t}$ for $r<<1$.
$Z(r,t)$ forgets the initial value $X(1)$ and
fluctuates around $\alpha/\theta$
with variance $B^2/(2\theta)$ for $r<<1$.
In the limit $r\to 0$, $D(r,t)\to t$, and $C(t)$ becomes constant.
The stochastic process remembers its initial condition $X(1)$ and
the process is not stationary. A discontinuous phase
transition occurs at $r=0$.

\subsection{Power-law decay memory kernel model : $d(t)=(t+1)^{-\gamma}$}

We denote $D_{d}(t),S_{d}(t),Z_{d}(t)$ for
$d(t)=(t+1)^{-\gamma}$ as
$D_{\gamma}(t)=\sum_{s=1}^{t}(t-s+1)^{-\gamma}$,
$S_{\gamma}(t)=\sum_{s=1}^{t}X(s)(t-s+1)^{-\gamma}$, 
$Z_{\gamma}(t)=S_{\gamma}(t)/D_{\gamma}(t)$, respectively.
Different from the exponential decay memory kernel model,
it is impossible to decompose
$Z_{\gamma}(t+1)$ with $Z_{\gamma}(t)$ and $X(t+1)$ directly.
We introduce an auxiliary
field $\{Z(r,t),0<r<\infty\}$
and express $Z_{\gamma}(t)$ as the superposition of the field.
Using the decomposition of $Z(r,t+1)$ with $Z(r,t)$ and $X(t+1)$ of
(\ref{eq:Z_r}), we derive the SDE for the auxiliary field $Z(r,t)$ and
estimate $Z_{\gamma}(t)$. 

First, we express $(t+1)^{-\gamma}$ as the superposition of $e^{-rt}$
with the gamma distribution $f_{\gamma}(r)=r^{\gamma-1}e^{-r}/\Gamma(\gamma)$ as
\begin{equation}
(t+1)^{-\gamma}=\int_{0}^{\infty} e^{-rt}\cdot f_{\gamma}(r)dr
\label{eq:id}.
\end{equation}
We also express $D_{\gamma}(t)$ and $S_{\gamma}(t)$ as the superposition
of $D(r,t)$ and $S(r,t)$ with $f_{\gamma}(r)$.
\begin{eqnarray}
D_{\gamma}(t)&=&\int_{0}^{\infty} D(r,t)f_{\gamma}(r)dr  \nonumber \\
S_{\gamma}(t)&=&\int_{0}^{\infty} S(r,t)f_{\gamma}(r)dr 
=\int_{0}^{\infty}D(r,t)Z(r,t)f_{\gamma}(r)dr  \nonumber 
\end{eqnarray}
$Z_{\gamma}(t)$ is then written as
\[
Z_{\gamma}(t)=\frac{S_{\gamma}(t)}{D_{\gamma}(t)}
=\frac{\int_{0}^{\infty}Z(r,t)D(r,t)f_{\gamma}(r)dr}
{\int_{0}^{\infty}D(r,t)f_{\gamma}(r)dr}.
\]
We denote the average of function $A(r,t)$
with the weight $D(r,t)f_{\gamma}(r)$ by $<A(r,t)>$,
\[
<A(r,t)>\equiv \frac{\int_{0}^{\infty}A(r,t)D(r,t)f_{\gamma}(r)dr}{D_{\gamma}(t)}.
\]
The denominator ensures the identity $<1>=1$ and 
$Z_{\gamma}(t)$ is written as $<Z(r,t)>$.

Next we derive the SDE for the auxiliary field $\{Z(r,t),0<r<\infty\}$.
When $\{Z(r,t)=z(r),0<r<\infty\}$,
$Z_{\gamma}(t)$ is estimated as $Z_{\gamma}(t)=z_{\gamma}\equiv <z(r)>$.
The probability that $X(t+1)$ takes the value 1 with the condition
$\{Z(r,t)=z(r),0<r<\infty\}$ is then given as
\[
P(X(t+1)=1|\{Z(r,t)=z(r),0<r<\infty\})
=\frac{\alpha+D_{\gamma}(t)<z(r)>}{\theta+D_{\gamma}(t)}
=\frac{\alpha+D_{\gamma}(t)z_{\gamma}}{\theta+D_{\gamma}(t)}.
\]
As $Z(r,t+1)=Z(r,t)+\frac{X(t+1)-Z(r,t)}{D(r,t+1)}$ in (\ref{eq:Z_r}),
we have, 
\begin{eqnarray}
&&E(Z(r,t+1)|\{Z(r,t)=z(r),0<r<\infty\})
= z(r)+\frac{E(X(t+1)|Z_{\gamma}(t)=z_{\gamma})-z(r)}{D(r,t+1)} \nonumber \\
&=& z(r)+\frac{\theta z_{0}+D_{\gamma}(t)<z(r)>-(\theta+D_{\gamma}(t))z(r)}
{(\theta+D_{\gamma}(t))D(r,t+1)}. 
\end{eqnarray}
Here, we define $z_{0}=\alpha/\theta$.
We denote the weighted average of $z_{0}$ and $z_{\gamma}(t)=<z(r,t)>$
with the weights
$\theta$ and $D_{\gamma}(t)$ as $z(t)$,
\begin{equation}
z(t)\equiv\frac{\theta z_0+D_{\gamma}(t)z_{\gamma}(t)}{\theta+D_{\gamma}(t)}
=\frac{\theta z_0+D_{\gamma}(t)<z(r,t)>}{\theta+D_{\gamma}(t)} \label{eq:z_t}.
\end{equation}
Then, the conditional expected value is written as
\[
E(Z(r,t+1)|\{Z(r,t)=z(r),0<r<\infty \})
=z(r)+\frac{z(t)-z(r)}{D_{r}(t+1)}.
\]
Likewise, we estimate the conditional variance as
\begin{eqnarray}  
&&V(Z(r,t+1)|\{Z_{r}(t)=z_{r},0<r<\infty\})=
  \frac{V(X(t+1)|\{Z(r,t)=z(r),0<r<\infty\})}{D_{r}(t+1)^2}\nonumber \\
&=&\frac{(D_{\gamma}(t)z_{\gamma}+\theta z_0)(D_{\gamma}(t)(1-z_{\gamma})+\theta(1-z_0))}{D_{r}(t+1)^2(D_{\gamma}(t+1)+\theta)^2}=\frac{z(t)(1-z(t))}{D_{r}(t+1)^2}.
\nonumber 
\end{eqnarray}
We assume that z(t) does not converge to 0 nor 1 and
the numerator of the variance does not vanish.
We can replace the numerator of the variance  with
constant $B^2$ for simplicity without affecting the essence of the process. 
One reads the drift and the diffusion terms and 
the SDE for the auxiliary field $\{Z(r,t),0<r<\infty\}$ becomes
\begin{equation}
dZ(r,t)=\frac{Z(t)-Z(r,t)}{D_{r}(t)}+\frac{B}{D_{r}(t)}dW_t,0<r<\infty.
\label{eq:SDE_power}.
\end{equation}  
We note that the auxiliary field $\{Z(r,t),0<r<\infty \}$
shares the common Wiener process $W_t$ as it emerges from $X(t+1)$.

The "formal" solution for the auxiliary field $\{Z(r,t)\}$
to the initial value problem  $\{Z(r,1)=X(1)=x,0<r<\infty\}$ is
\[
Z(r,t)=I(r,t)\cdot x+I(r,t)\int_{1}^{t}\frac{Z(s)}{D(r,s)}I(r,s)^{-1}ds
+I(r,t)\int_{0}^{t}\frac{B}{D(r,s)}I(r,s)^{-1}dW_s,0<r<\infty. 
\]
Here, $I(r,t)$ is defined as
\[
I(r,t)\equiv \exp\left(-\int_{1}^{t}\frac{1}{D(r,s)}ds\right)
=\left(\frac{e^{rt}-1}{e^r-1}\right)^{-\frac{1-e^{-r}}{r}}.
\]
The reason to call the solution "formal" is that $Z(s)$ in the integral
of the second term of the solution
is calculated using $\{Z(r,t),0<r<\infty\}$.
We consider the deviation of $Z(r,t)$ from $z_{0}$
as $\Delta Z(r,t)$, i.e., $Z(r,t)=z_{0}+\Delta Z(r,t)$.
$\Delta Z_{\gamma}(t)$ is also defined as $\Delta Z_{\gamma}(t)=<\Delta Z(r,t)>
=<Z(r,t)-z_{0}>=Z_{\gamma}(t)-z_{0}$.
$Z(t)$ is then written as
\[
Z(t)=z_{0}+\frac{D_{\gamma}(t)\Delta Z_{\gamma}(t)}{\theta+D_{\gamma}(t)}.
\]
The formal solution for $\{Z(r,t)\}$ becomes 
\begin{eqnarray}
Z(r,t)&=&I(r,t)\cdot x+(1-I(r,t))\cdot z_{0} \nonumber \\
&+& I(r,t)
\left(\int_{1}^{t}\frac{1}{D(r,s)}\left(\frac{D_{\gamma}(s)\Delta z_{\gamma}(s)}{\theta+D_{\gamma}(s)}\right)I(r,s)^{-1}ds
+\int_{0}^{t}\frac{B}{D(r,s)}I(r,s)^{-1}dW_s\right),0<r<\infty .
\nonumber
\end{eqnarray}
$Z_{\gamma}(t)=<Z(r,t)>$ is then estimated as
\begin{eqnarray}
Z_{\gamma}(t)&=&z_{0}+<I_{r}(t)>(x-z_{0}) \nonumber \\
&+&\left<I(r,t)
(\int_{1}^{t}\frac{1}{D(r,s)}\left(\frac{D_{\gamma}(s)\Delta Z_{\gamma}(s)}{\theta+D_{\gamma}(s)}\right)I(r,s)^{-1}ds
+\int_{0}^{t}\frac{B}{D(r,s)}I(r,s)^{-1}dW_s)\right>\nonumber .
\end{eqnarray}

To estimate the average $<>$ in the right hand side of the solution
for $Z_{\gamma}(t)$, we derive a useful approximation integral formula
for the arbitrary function $f(r)$, which is valid for large $t$.
\[
\int_{0}^{\infty}f(r) f_{\gamma}(r)e^{-rt}dr=
\int_{0}^{1/(t+1)}f(r)\cdot \gamma r^{\gamma-1}dr.
\]
It replaces the definite integral of $f(r)$
multiplied by $f_{\gamma}(r)e^{-rt}$ over $r\in [0,\infty)$
  with the definite integral of $f(r)$ multiplied by $\gamma r^{\gamma-1}$
  over $r\in [0,1/(t+1)]$.
For large $t$, $e^{-rt}<<1$ for $r>>1/t$
and $f(r)$ with $r\sim 0$ dominates the asymptotic behavior.
In the limit $r\to 0$, $f_{\gamma}(r)$ can be replaced with $\gamma r^{\gamma-1}$.
Here, we adopt $\gamma r^{\gamma-1}$ instead of $r^{\gamma-1}/\Gamma(\gamma)$
to ensure that the approximation formula gives $(t+1)^{-\gamma}$
for $f(r)=1$, which is the identity of (\ref{eq:id}).
The formula is useful to study the asymptotic (large $t$)
 behavior of the left-hand side of the equation.

Then, we have two approximation formulas,
\begin{eqnarray}
<I_{r}(t)>&\simeq&
\frac{t^{-\gamma}}{D_{\gamma}(t)},\nonumber \\
<I_{r}(t)\frac{1}{D_{r}(s)}I_{r}(s)^{-1}>&\simeq &
\frac{(t-s+1)^{-\gamma}}{D_{\gamma}(t)} \nonumber.
\end{eqnarray}
Using the results, we have
\begin{eqnarray}
  Z_{\gamma}(t)&=&z_{0}+\frac{t^{-\gamma}}{D_{\gamma}(t)}(x-z_{0}) \nonumber \\
  &+&
\frac{1}{D_{\gamma}(t)}\left(
\int_{1}^{t}(t-s+1)^{-\gamma}\left(\frac{D_{\gamma}(s)\Delta z_{\gamma}(s)}{\theta+D_{\gamma}(s)}\right)ds+\int_{1}^{t}(t-s+1)^{-\gamma}BdW_s\right).\nonumber 
\end{eqnarray}
The conditional expected value of $Z_{\gamma}(t)$ for $X(1)=x$ is
\begin{eqnarray}
E(Z_{\gamma}(t)|X(1)=x)&=&
z_{0}+\frac{t^{-\gamma}}{D_{\gamma}(t)}(x-z_{0}) \nonumber \\
&+&
\frac{1}{D_{\gamma}(t)}\left(
\int_{1}^{t}(t-s+1)^{-\gamma}\left(\frac{D_{\gamma}(s)(E(Z_{\gamma}(s)|X(1)=x)-z_{0})}{\theta+D_{\gamma}(s)}\right)ds\right)\nonumber .
\end{eqnarray}
$C(t)$ is then given as
\begin{equation}
C(t)=\frac{1}{D_{\gamma}(t)+\theta}\left(t^{-\gamma}+\int_{1}^{t}(t-s+1)^{-\gamma}C(s)\right)   \label{eq:self_power}.
\end{equation}
We derive the integral equation for $C(t)$, which is
the continuous approximation of (\ref{eq:self}).

\section{\label{sec:results}Asymptotic behavior of Correlation function}

We study the asymptotic behavior of the ACF for the power-law decay memory kernel
model. We assume the power-law asymptotic behavior of the leading and sub-leading terms of the ACF as
\begin{equation}
  C(t)=ct^{-\delta}+c't^{-\delta'},\delta'>\delta\,\, \mbox{for}\,\,
  t>>1 \label{eq:asymptotic}.
\end{equation}
We multiply $(D_{\gamma}(t)+\theta)$ on both sides of (\ref{eq:self_power})
and we obtain
\begin{equation}
  (D_{\gamma}(t)+\theta)C(t)=\left(t^{-\gamma}+\int_{1}^{t}(t-s+1)^{-\gamma}C(s)ds\right).
  \label{eq:self2}
\end{equation}
We substitute (\ref{eq:asymptotic}) into (\ref{eq:self_power}) and
compare the power-law exponent of the leading and subleading terms
on both sides of the equation.

The asymptotic behaviors of
$D_{\gamma}(t)$ are $1/(\gamma-1),\ln t$ and $t^{1-\gamma}/(1-\gamma)$
for $\gamma>1,=1$ and $<1$, respectively.
We put $ct^{-\delta}+c't^{-\delta'}$ in $C(t)$ and
estimate the integral of the right hand side of (\ref{eq:self2}).
\[
\int_{1}^{t}(t-s+1)^{-\gamma}(cs^{-\delta}+cs^{-\delta'})ds.
\]
After changing the variable from $s$ to $x=s/(t+1)$, the first term
of the integral becomes 
\[
c(t+1)^{1-\delta-\gamma}\int_{\epsilon}^{1-\epsilon}(1-x)^{-\gamma}x^{-\delta}dx.
\]
Here we write $\epsilon=1/(t+1)$.
If the exponents $\gamma,\delta$ of $x,1-x$
in the integrand are smaller than 1,
the integral converges to the beta function, $B(1-\gamma,1-\delta)$,
in the limit $\epsilon\to 0$.
If $\gamma\ge 1$ or $\delta \ge 1$, it is necessary to
regularize the integral.
We denote the integral as $B(1-\gamma,1-\delta,\epsilon)$,
\[
B(1-\gamma,1-\delta,\epsilon)\equiv \int_{\epsilon}^{1-\epsilon}(1-x)^{-\gamma}x^{-\delta}dx.
\]
We write the finite term of $B(1-\gamma,1-\delta,\epsilon)$ in the limit
$\epsilon$ as
$B_{reg}(1-\gamma,1-\delta)$.
\[
B(1-\gamma,1-\delta,\epsilon)=B_{reg}(1-\gamma,1-\delta)+\mbox{divergent terms}.
\]
When $\gamma<1,\delta<1$, the integral converges and 
$B_{reg}(1-\gamma,1-\delta)$ is $B(1-\gamma,1-\delta)$.

The regularization procedure is simple.
When $\gamma>1$ and $\delta <1$, the integral of $(1-x)^{-\gamma}x^{-\delta}$
diverges when the upper bound $1-\epsilon$ goes to 1.
We regularize the integral as
\begin{eqnarray}
B(1-\gamma,1-\delta,\epsilon)&=&\int_{1/\epsilon}^{1-\epsilon}(1-x)^{-\gamma}(x^{-\delta}-1)dx
+\int_{1/\epsilon}^{1-\epsilon}(1-x)^{-\gamma}dx \nonumber \\
&=&B_{reg}(1-\gamma,1-\delta,\epsilon)+\frac{1}{\gamma-1}\epsilon^{1-\gamma}.
\nonumber 
\end{eqnarray}
Here we neglect a term that vanishes in the limit $\epsilon\to 0$.
The second term diverges as $\epsilon^{1-\gamma}\propto t^{\gamma-1}$.
Likewise for $\gamma>1,\delta>1$, the divergence of 
the integral of $(1-x)^{-\gamma}x^{-\delta}$ comes from both bounds
of the integral.
We regularize the integral as
\begin{eqnarray}
B(1-\gamma,1-\delta,\epsilon)&=&\int_{1/\epsilon}^{1-\epsilon}(((1-x)^{-\gamma}-1)(x^{-delta}-1)-1)dx
+\int_{1/\epsilon}^{1-\epsilon}(1-x)^{-\gamma}dx+\int_{1/\epsilon}^{1-\epsilon}x^{-\delta}dx \nonumber \\
&=&B_{reg}(1-\gamma,1-\delta,\epsilon)+\frac{1}{\gamma-1}\epsilon^{1-\gamma}+\frac{1}{\delta-1}\epsilon^{1-\delta}.
\nonumber 
\end{eqnarray}
When $\gamma=1$ and $\delta<1$, a logarithmic divergence appears, and we have
\begin{eqnarray}
B(0,1-\delta,\epsilon)&=&\int_{1/\epsilon}^{1-\epsilon}((1-x)^{-1})(x^{-delta}-1)dx
+\int_{1/\epsilon}^{1-\epsilon}(1-x)^{-1}dx\nonumber \\
&=&B_{reg}(0,1-\delta,\epsilon)-\ln \epsilon.
\nonumber 
\end{eqnarray}
Using $B(1-\gamma,1-\delta),B(1-\gamma,1-\delta')$,
(\ref{eq:self2}) becomes
\begin{equation}
(D_{\gamma}(t)+\theta)(ct^{-\delta}+c't^{-\delta'})=t^{-\gamma}
  +t^{1-\gamma-\delta}B(1-\gamma,1-\delta,\epsilon)+t^{1-\gamma-\delta'}B(1-\gamma,1-\delta',\epsilon).
\label{eq:self3}
\end{equation}

\begin{enumerate}
\item $\gamma<1$ case.

$D_{\gamma}(t)=t^{1-\gamma}/(1-\gamma)$ and we assume $\delta<1$.
(\ref{eq:self3}) becomes,
\begin{eqnarray}
(\theta+t^{1-\gamma}/(1-\gamma))(ct^{-\delta}+
c't^{-\delta'})&=&t^{-\gamma}+
ct^{1-\gamma-\delta}B(1-\gamma,1-\delta) \nonumber \\
&+&c't^{1-\gamma-\delta'}(
B_{reg}(1-\gamma,1-\delta')+\frac{1}{\delta'-1}t^{\delta'-1})\nonumber .  
\end{eqnarray}
The last term $\frac{1}{\delta'-1}t^{\delta'-1}$ does not appear
when $\delta'<1$.
The leading term of the left-hand side is $c/(1-\gamma)t^{1-\gamma-\delta}$.
The leading term of the right-hand side
is $c B(1-\gamma,1-\delta)t^{1-\gamma-\delta}$
as $1-\gamma-\delta>1-\gamma-\delta'$ and $1-\gamma-\delta>-\gamma$.
By equating the 
coefficients of the leading terms, we obtain
an identity
\[
\frac{1}{1-\gamma}=B(1-\gamma,1-\delta).
\]
As $B(1-\gamma,1)=1/(1-\gamma)$, we obtain the result $\delta=0$.
The subleading term of the left-hand side is $O(t^0)$ as $\delta=0$, we
have $1-\gamma-\delta'=0$. We obtain $\delta'=1-\gamma$. 
If one assume $\delta>1$, we cannot match $O(t^0)$
terms of both sides of (\ref{eq:self3}). We can exclude the
possibility $\delta,\delta'>1$ for $\gamma<1$.

\item $\gamma=1$ case.

$D_{\gamma}(t)=\ln t$ and  we assume $\delta<1$.
(\ref{eq:self3}) becomes,
\[
(\theta+\ln t)(ct^{-\delta}+c't^{-\delta'})=t^{-1}+
ct^{-\delta}(B_{reg}(0,1-\delta)+\ln t)+c't^{-\delta'}(
B_{reg}(0,1-\delta')+\ln t+\frac{1}{\delta'-1}t^{\delta'-1}).  
\]
The leading term is $O(\ln t\cdot t^{-\delta})$ and we obtain a trivial
identity $c=c$.
The subleading term is $O(t^{-\delta})$ and
we obtain the identity
\begin{equation}
\theta=B_{reg}(0,1-\delta).
\end{equation}
By solving the identity, we estimate $\delta_{c}$, the critical
exponent of the ACF, $C(t)\propto t^{-\delta_c}$.
The estimation of $\delta'$ is obscure, as t
The same identity with $\delta$ holds for $\delta'$
by comparing  the  $O(t^{-\delta'})$ terms. It suggests
that $\delta'=\delta$ at $\gamma=1$.

\item $\gamma>1$ case.

$D_{\gamma}(t)=1/(\gamma-1)$ and we assume $1<\delta<\delta'$.
(\ref{eq:self3}) becomes,
\begin{eqnarray}
(\theta+1/(\gamma-1))(ct^{-\delta}+c't^{-\delta'})&=&t^{-1}+
ct^{1-\gamma-\delta}(B_{reg}(1-\gamma,1-\delta)+\frac{1}{\gamma-1}t^{\gamma-1}+\frac{1}{\delta-1}t^{\delta-1}) \nonumber \\
&+&c't^{1-\gamma-\delta'}
(B_{reg}(1-\gamma,1-\delta')+\frac{1}{\gamma-1}t^{\gamma-1}+\frac{1}{\delta'-1}t^{\delta'-1})\nonumber .  
\end{eqnarray}
The leading term is $O(t^{-\delta})$ and we obtain $\delta=\gamma$.
The subleading term is $O(t^{-\delta'})$ and we obtain $\delta'=2\gamma-1$.

\end{enumerate}

The power-law exponents of the leading and subleading terms of 
the ACF $C(t)$ are summarized as
\begin{eqnarray}
 (\delta,\delta') =\left\{
    \begin{array}{cc}
      (\gamma,2\gamma-1) & \gamma>1 \\
      (\delta_c,\delta_c) & \gamma=1 \\
      (0,1-\gamma)      & \gamma<1
    \end{array}
    \right. \label{eq:delta_th}
\end{eqnarray}
We denote the values by $(\delta_{th},\delta_{th}')$ to distinguish them
from the numerically estimated values,
which are denoted as $(\delta_{num},\delta_{num}')$.

We solve the recursive relation of 
(\ref{eq:self}) numerically and estimate $C(t)$ for $t\le 2^8\times 10^4$.
We set $(\alpha,\beta)=(1,1)$ and $\gamma \in [0.0,0.1,0,2,\cdots,2.0]$.
For large $t>>1$, the leading term of $C(t)$ is $ct^{-\delta}$, and
we have the relation for $\delta$,  
\begin{equation}
\delta=\lim_{t\to\infty}\ln_{2}C(t)/C(2t) \label{eq:delta}.
\end{equation}
We estimate $\delta_{num}$ by the formula with $t=2^7\times 10^4$.
About $\delta_{num}'$, we use the relation 
\begin{equation}
\delta'=\lim_{t\to\infty}
\ln_{2}\frac{C(t)-C(2t)2^{\delta}}{C(2t)-C(4t)2^{\delta}} \label{eq:delta_d}.
\end{equation}
The relation is based on the relation for the subleading term
of $C(t)$, $C(t)-C(2t)2^{\delta}\simeq c't^{-\delta'}(1-2^{\delta-\delta'})$
for $t>>1$. We estimate $\delta_{num}'$ using the formula $t=2^6\times 10^4$.

\begin{figure}[h]
\begin{center}
\includegraphics[clip, width=12cm]{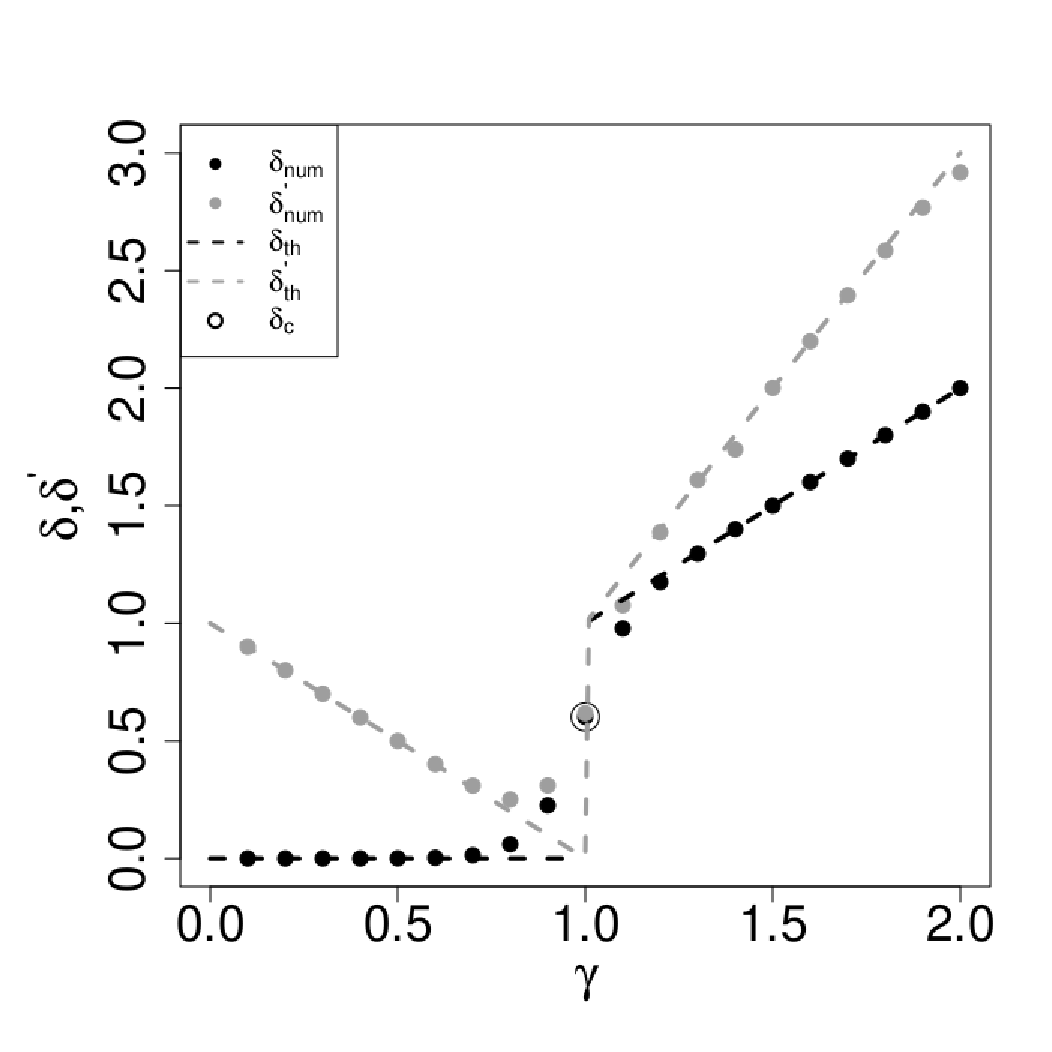}
\caption{Plot of $(\delta_{num},\delta_{num}')$ vs. $\gamma$.
  We estimate the power-law exponents using (\ref{eq:delta}) and (\ref{eq:delta_d}). $(\delta_{th},\delta_{th}')$ are the theoretically predicted values
in (\ref{eq:delta_th}).}
\label{fig:gamma_vs_delta}
\end{center}
\end{figure}

Figure \ref{fig:gamma_vs_delta} shows $(\delta_{num},\delta_{num}')$
vs. $\gamma$. For reference, we also plot
$(\delta_{th},\delta_{th}')$. As can be clearly observed, if 
$\gamma$ is far from the critical value $\gamma_{c}=1$, the
numerically estimated
values agree with the theoretical
results. However, near the critical point, the discrepancy becomes
large. At the critical point, both $\delta_{num},\delta_{num}'$ coincide with
$\delta_c$.

\section{\label{sec:conclusion}Conclusion}

We study the P\'{o}lya urn with an arbitrary memory kernel.
Previously, we have shown that the process shows a
phase transition for the power-law decay memory kernel\cite{Hisakado:2020}.
We formulate the problem using the method of stochastic approximation.
It is necessary to introduce an auxiliary field $\{Z(r,t),0<r<\infty\}$
to reduce the process Markovian. The SDE for the auxiliary field
are coupled as the field
is driven by a common Wiener process.
We obtain the formal solution for the initial value problem and
derive the self-consistent equation
for the ACF in the framework. We estimate the power-law exponents
for the leading and subleading terms of the ACF. The results for
the power-law exponents are new except for $\delta_c$
at the critical point.  

The essential difference from the continuous phase transitions of
the nonlinear P\'{o}lya urn is the lack of the universal function.
The transition of nonlinear P\'{o}lya urns
has the correlation length scale $\xi$ for the decay of the ACF
and the asymptotic behavior of the ACF obeys a scaling relation.
For the existence of the scaling relation for ACF, the continuity of
the power-law exponent and the logarithmic behavior of the ACF
at the critical point is crucial\cite{Mori:2015-2,Nakayama:2021}.
In the phase transition of the P\'{o}lya urn with power-law decay,
the power-law exponent is discontinuous at $\gamma=1$
and the ACF does not show logarithmic
behavior at the critical point.
The phase transition of the P\'{o}lya urn with power-law
memory kernel is completely different from that
of the nonlinear P\'{o}lya urn.
The nature of the former
is the phase transition of the stationary-nonstationary phase transition.
The nature of the latter is the absorption-state phase transition
where there exists the scaling relation for the order parameter.

This study also shows the importance of the correlation of the Wiener
process. The auxiliary field $\{Z(r,t),0<r<\infty\}$, which is introduced for
the P\'{o}lya urn with power-law decay memory kernel, 
shares the same Wiener
process $W_t$ and it induces the coupling among the modes of the field.
If $\{Z(r,t),0<r<\infty\}$ are driven by independent Wiener processes $\{W(r,t),0<r<\infty\}$,
they obey the Ornstein-Uhlenbeck process for the exponential decay memory kernel
independently from each other. They are mean-reverting and 
converge to the same equilibrium $\alpha/\theta$.
Even though $Z(r,t)$ with small $r$ should have long relaxation time,
$Z_{\gamma}(t)=<Z(r,t)>$ also converges to the same equilibrium.
The phase transition does not occur in the case.
The condition of
the mean-revering for the multivariate Ornstein-Uhlenbeck
process is known\cite{Vatiutipong:2019}.
By the correlation of the Wiener process, the condition
should be broken.
To understand  the importance of the correlation of the Wiener
process for the asymptotic behavior of the process, we need to
solve the initial value problem for the SDE of the coupled
auxiliary field.

\bibliography{myref202009}

\providecommand{\noopsort}[1]{}\providecommand{\singleletter}[1]{#1}%
\begin{thebibliography}{24}%
\makeatletter
\providecommand \@ifxundefined [1]{%
 \@ifx{#1\undefined}
}%
\providecommand \@ifnum [1]{%
 \ifnum #1\expandafter \@firstoftwo
 \else \expandafter \@secondoftwo
 \fi
}%
\providecommand \@ifx [1]{%
 \ifx #1\expandafter \@firstoftwo
 \else \expandafter \@secondoftwo
 \fi
}%
\providecommand \natexlab [1]{#1}%
\providecommand \enquote  [1]{``#1''}%
\providecommand \bibnamefont  [1]{#1}%
\providecommand \bibfnamefont [1]{#1}%
\providecommand \citenamefont [1]{#1}%
\providecommand \href@noop [0]{\@secondoftwo}%
\providecommand \href [0]{\begingroup \@sanitize@url \@href}%
\providecommand \@href[1]{\@@startlink{#1}\@@href}%
\providecommand \@@href[1]{\endgroup#1\@@endlink}%
\providecommand \@sanitize@url [0]{\catcode `\\12\catcode `\$12\catcode
  `\&12\catcode `\#12\catcode `\^12\catcode `\_12\catcode `\%12\relax}%
\providecommand \@@startlink[1]{}%
\providecommand \@@endlink[0]{}%
\providecommand \url  [0]{\begingroup\@sanitize@url \@url }%
\providecommand \@url [1]{\endgroup\@href {#1}{\urlprefix }}%
\providecommand \urlprefix  [0]{URL }%
\providecommand \Eprint [0]{\href }%
\providecommand \doibase [0]{http://dx.doi.org/}%
\providecommand \selectlanguage [0]{\@gobble}%
\providecommand \bibinfo  [0]{\@secondoftwo}%
\providecommand \bibfield  [0]{\@secondoftwo}%
\providecommand \translation [1]{[#1]}%
\providecommand \BibitemOpen [0]{}%
\providecommand \bibitemStop [0]{}%
\providecommand \bibitemNoStop [0]{.\EOS\space}%
\providecommand \EOS [0]{\spacefactor3000\relax}%
\providecommand \BibitemShut  [1]{\csname bibitem#1\endcsname}%
\let\auto@bib@innerbib\@empty
\bibitem [{\citenamefont {Mantegna}\ and\ \citenamefont
  {Stanley}(2007)}]{Mantegna:2008}%
  \BibitemOpen
  \bibfield  {author} {\bibinfo {author} {\bibfnamefont {R.~N.}\ \bibnamefont
  {Mantegna}}\ and\ \bibinfo {author} {\bibfnamefont {H.~E.}\ \bibnamefont
  {Stanley}},\ }\href@noop {} {\emph {\bibinfo {title} {Introduction to
  Econophysics: Correlations and Complexity in Finance}}}\ (\bibinfo
  {publisher} {Cambridge University Press, Cambridge},\ \bibinfo {year}
  {2007})\BibitemShut {NoStop}%
\bibitem [{\citenamefont {Galam}(2008)}]{Galam:2008}%
  \BibitemOpen
  \bibfield  {author} {\bibinfo {author} {\bibfnamefont {S.}~\bibnamefont
  {Galam}},\ }\href@noop {} {\bibfield  {journal} {\bibinfo  {journal} {Int. J.
  Mod. Phys. C}\ }\textbf {\bibinfo {volume} {19}},\ \bibinfo {pages} {409}
  (\bibinfo {year} {2008})}\BibitemShut {NoStop}%
\bibitem [{\citenamefont {Castellano}\ \emph {et~al.}(2009)\citenamefont
  {Castellano}, \citenamefont {Fortunato},\ and\ \citenamefont
  {Loreto}}]{Castellano:2009}%
  \BibitemOpen
  \bibfield  {author} {\bibinfo {author} {\bibfnamefont {C.}~\bibnamefont
  {Castellano}}, \bibinfo {author} {\bibfnamefont {S.}~\bibnamefont
  {Fortunato}}, \ and\ \bibinfo {author} {\bibfnamefont {V.}~\bibnamefont
  {Loreto}},\ }\href@noop {} {\bibfield  {journal} {\bibinfo  {journal}
  {Rev.Mod.Phys.}\ }\textbf {\bibinfo {volume} {81}},\ \bibinfo {pages} {591}
  (\bibinfo {year} {2009})}\BibitemShut {NoStop}%
\bibitem [{\citenamefont {Arthur}(1989)}]{Arthur:1989}%
  \BibitemOpen
  \bibfield  {author} {\bibinfo {author} {\bibfnamefont {W.~B.}\ \bibnamefont
  {Arthur}},\ }\href@noop {} {\bibfield  {journal} {\bibinfo  {journal} {Econ.
  Jour.}\ }\textbf {\bibinfo {volume} {99}},\ \bibinfo {pages} {116} (\bibinfo
  {year} {1989})}\BibitemShut {NoStop}%
\bibitem [{\citenamefont {Bikhchandani}\ \emph {et~al.}(1992)\citenamefont
  {Bikhchandani}, \citenamefont {Hirshleifer},\ and\ \citenamefont
  {Welch}}]{Bikchandani:1992}%
  \BibitemOpen
  \bibfield  {author} {\bibinfo {author} {\bibfnamefont {S.}~\bibnamefont
  {Bikhchandani}}, \bibinfo {author} {\bibfnamefont {D.}~\bibnamefont
  {Hirshleifer}}, \ and\ \bibinfo {author} {\bibfnamefont {I.}~\bibnamefont
  {Welch}},\ }\href@noop {} {\bibfield  {journal} {\bibinfo  {journal} {J.
  Polit. Econ.}\ }\textbf {\bibinfo {volume} {100}},\ \bibinfo {pages} {992}
  (\bibinfo {year} {1992})}\BibitemShut {NoStop}%
\bibitem [{\citenamefont {Cont}\ and\ \citenamefont
  {Bouchaud}(2000)}]{Cont:2000}%
  \BibitemOpen
  \bibfield  {author} {\bibinfo {author} {\bibfnamefont {R.}~\bibnamefont
  {Cont}}\ and\ \bibinfo {author} {\bibfnamefont {J.}~\bibnamefont
  {Bouchaud}},\ }\href@noop {} {\bibfield  {journal} {\bibinfo  {journal}
  {Macroecon. Dynam.}\ }\textbf {\bibinfo {volume} {4}},\ \bibinfo {pages}
  {170} (\bibinfo {year} {2000})}\BibitemShut {NoStop}%
\bibitem [{\citenamefont {Mori}\ \emph {et~al.}(2016)\citenamefont {Mori},
  \citenamefont {Nakayama},\ and\ \citenamefont {Hisakado}}]{Mori:2016}%
  \BibitemOpen
  \bibfield  {author} {\bibinfo {author} {\bibfnamefont {S.}~\bibnamefont
  {Mori}}, \bibinfo {author} {\bibfnamefont {K.}~\bibnamefont {Nakayama}}, \
  and\ \bibinfo {author} {\bibfnamefont {M.}~\bibnamefont {Hisakado}},\
  }\href@noop {} {\bibfield  {journal} {\bibinfo  {journal} {Phys.Rev. E}\
  }\textbf {\bibinfo {volume} {94}},\ \bibinfo {pages} {052301} (\bibinfo
  {year} {2016})}\BibitemShut {NoStop}%
\bibitem [{\citenamefont {Kirman}(1993)}]{Kirman:1993}%
  \BibitemOpen
  \bibfield  {author} {\bibinfo {author} {\bibfnamefont {A.}~\bibnamefont
  {Kirman}},\ }\href@noop {} {\bibfield  {journal} {\bibinfo  {journal} {Q. J.
  Econ.}\ }\textbf {\bibinfo {volume} {108}},\ \bibinfo {pages} {137} (\bibinfo
  {year} {1993})}\BibitemShut {NoStop}%
\bibitem [{\citenamefont {Lux}(1995)}]{Lux:1995}%
  \BibitemOpen
  \bibfield  {author} {\bibinfo {author} {\bibfnamefont {T.}~\bibnamefont
  {Lux}},\ }\href@noop {} {\bibfield  {journal} {\bibinfo  {journal} {Econ.
  J.}\ }\textbf {\bibinfo {volume} {105}},\ \bibinfo {pages} {881} (\bibinfo
  {year} {1995})}\BibitemShut {NoStop}%
\bibitem [{\citenamefont {Hisakado}\ and\ \citenamefont
  {Mori}(2020)}]{Hisakado:2020}%
  \BibitemOpen
  \bibfield  {author} {\bibinfo {author} {\bibfnamefont {M.}~\bibnamefont
  {Hisakado}}\ and\ \bibinfo {author} {\bibfnamefont {S.}~\bibnamefont
  {Mori}},\ }\href@noop {} {\bibfield  {journal} {\bibinfo  {journal} {Physica
  A}\ }\textbf {\bibinfo {volume} {544}},\ \bibinfo {pages} {123480} (\bibinfo
  {year} {2020})}\BibitemShut {NoStop}%
\bibitem [{\citenamefont {Pluto}\ and\ \citenamefont
  {Tasche}(2011)}]{Pluto:2011}%
  \BibitemOpen
  \bibfield  {author} {\bibinfo {author} {\bibfnamefont {K.}~\bibnamefont
  {Pluto}}\ and\ \bibinfo {author} {\bibfnamefont {D.}~\bibnamefont {Tasche}},\
  }in\ \href@noop {} {\emph {\bibinfo {booktitle} {The Basel II Risk
  Parameters}}},\ \bibinfo {editor} {edited by\ \bibinfo {editor} {\bibnamefont
  {B.Engelmann.}}\ and\ \bibinfo {editor} {\bibfnamefont {R.}~\bibnamefont
  {Rauhmeier}}}\ (\bibinfo  {publisher} {Springer, Berlin},\ \bibinfo {year}
  {2011})\ pp.\ \bibinfo {pages} {75--101}\BibitemShut {NoStop}%
\bibitem [{\citenamefont {Hisakado}\ and\ \citenamefont
  {Mori}(2021)}]{Hisakado:2021}%
  \BibitemOpen
  \bibfield  {author} {\bibinfo {author} {\bibfnamefont {M.}~\bibnamefont
  {Hisakado}}\ and\ \bibinfo {author} {\bibfnamefont {S.}~\bibnamefont
  {Mori}},\ }\href@noop {} {\bibfield  {journal} {\bibinfo  {journal} {Physica
  A}\ }\textbf {\bibinfo {volume} {563}},\ \bibinfo {pages} {125435} (\bibinfo
  {year} {2021})}\BibitemShut {NoStop}%
\bibitem [{\citenamefont {G.P\'{o}lya}(1931)}]{Polya:1931}%
  \BibitemOpen
  \bibfield  {author} {\bibinfo {author} {\bibnamefont {G.P\'{o}lya}},\
  }\href@noop {} {\bibfield  {journal} {\bibinfo  {journal} {Ann. Inst. Henri
  Poincar\'{e}}\ }\textbf {\bibinfo {volume} {1}},\ \bibinfo {pages} {117}
  (\bibinfo {year} {1931})}\BibitemShut {NoStop}%
\bibitem [{\citenamefont {Kanazawa}\ and\ \citenamefont
  {D.Sornette}(2020{\natexlab{a}})}]{Kanazawa:2020}%
  \BibitemOpen
  \bibfield  {author} {\bibinfo {author} {\bibfnamefont {K.}~\bibnamefont
  {Kanazawa}}\ and\ \bibinfo {author} {\bibnamefont {D.Sornette}},\ }\href@noop
  {} {\bibfield  {journal} {\bibinfo  {journal} {Phys. Rev. Lett.}\ }\textbf
  {\bibinfo {volume} {125}},\ \bibinfo {pages} {138301} (\bibinfo {year}
  {2020}{\natexlab{a}})}\BibitemShut {NoStop}%
\bibitem [{\citenamefont {Kanazawa}\ and\ \citenamefont
  {D.Sornette}(2020{\natexlab{b}})}]{Kanazawa:2020-2}%
  \BibitemOpen
  \bibfield  {author} {\bibinfo {author} {\bibfnamefont {K.}~\bibnamefont
  {Kanazawa}}\ and\ \bibinfo {author} {\bibnamefont {D.Sornette}},\ }\href@noop
  {} {\bibfield  {journal} {\bibinfo  {journal} {Phys. Rev. Research.}\
  }\textbf {\bibinfo {volume} {2}},\ \bibinfo {pages} {033442} (\bibinfo {year}
  {2020}{\natexlab{b}})}\BibitemShut {NoStop}%
\bibitem [{\citenamefont {Hawkes}(1971)}]{Hawkes:1971}%
  \BibitemOpen
  \bibfield  {author} {\bibinfo {author} {\bibfnamefont {A.}~\bibnamefont
  {Hawkes}},\ }\href@noop {} {\bibfield  {journal} {\bibinfo  {journal} {J. R.
  Stat. Soc. Ser. B}\ }\textbf {\bibinfo {volume} {33}},\ \bibinfo {pages}
  {438} (\bibinfo {year} {1971})}\BibitemShut {NoStop}%
\bibitem [{\citenamefont {Mori}\ and\ \citenamefont
  {Hisakado}(2015{\natexlab{a}})}]{Mori:2015-2}%
  \BibitemOpen
  \bibfield  {author} {\bibinfo {author} {\bibfnamefont {S.}~\bibnamefont
  {Mori}}\ and\ \bibinfo {author} {\bibfnamefont {M.}~\bibnamefont
  {Hisakado}},\ }\href@noop {} {\bibfield  {journal} {\bibinfo  {journal}
  {Phys.Rev. E}\ }\textbf {\bibinfo {volume} {92}},\ \bibinfo {pages} {052112}
  (\bibinfo {year} {2015}{\natexlab{a}})}\BibitemShut {NoStop}%
\bibitem [{\citenamefont {Nakayama}\ and\ \citenamefont
  {Mori}(2021)}]{Nakayama:2021}%
  \BibitemOpen
  \bibfield  {author} {\bibinfo {author} {\bibfnamefont {K.}~\bibnamefont
  {Nakayama}}\ and\ \bibinfo {author} {\bibfnamefont {S.}~\bibnamefont
  {Mori}},\ }\href@noop {} {\bibfield  {journal} {\bibinfo  {journal} {Phys.
  Rev.E}\ }\textbf {\bibinfo {volume} {104}},\ \bibinfo {pages} {014109}
  (\bibinfo {year} {2021})}\BibitemShut {NoStop}%
\bibitem [{\citenamefont {Hinrichsen}(2000)}]{Hinrichsen:2000}%
  \BibitemOpen
  \bibfield  {author} {\bibinfo {author} {\bibfnamefont {H.}~\bibnamefont
  {Hinrichsen}},\ }\href@noop {} {\bibfield  {journal} {\bibinfo  {journal}
  {Adv.Phys.}\ }\textbf {\bibinfo {volume} {49}},\ \bibinfo {pages} {815}
  (\bibinfo {year} {2000})}\BibitemShut {NoStop}%
\bibitem [{\citenamefont {Mori}\ and\ \citenamefont
  {Hisakado}(2015{\natexlab{b}})}]{Mori:2015}%
  \BibitemOpen
  \bibfield  {author} {\bibinfo {author} {\bibfnamefont {S.}~\bibnamefont
  {Mori}}\ and\ \bibinfo {author} {\bibfnamefont {M.}~\bibnamefont
  {Hisakado}},\ }\href@noop {} {\bibfield  {journal} {\bibinfo  {journal}
  {J.Phys.Soc.Jpn.}\ }\textbf {\bibinfo {volume} {84}},\ \bibinfo {pages}
  {054001} (\bibinfo {year} {2015}{\natexlab{b}})}\BibitemShut {NoStop}%
\bibitem [{\citenamefont {B$\ddot{o}$hm}(2000)}]{Bohm:2000}%
  \BibitemOpen
  \bibfield  {author} {\bibinfo {author} {\bibfnamefont {W.}~\bibnamefont
  {B$\ddot{o}$hm}},\ }\href@noop {} {\bibfield  {journal} {\bibinfo  {journal}
  {J. Appl. Prob.}\ }\textbf {\bibinfo {volume} {37}},\ \bibinfo {pages} {470}
  (\bibinfo {year} {2000})}\BibitemShut {NoStop}%
\bibitem [{\citenamefont {Gardiner}(2009)}]{Gardiner:2009}%
  \BibitemOpen
  \bibfield  {author} {\bibinfo {author} {\bibfnamefont {C.}~\bibnamefont
  {Gardiner}},\ }\href@noop {} {\emph {\bibinfo {title} {Stochastic Methods: A
  handbook for the Natural and Social Science, 4th ed.}}}\ (\bibinfo
  {publisher} {Springer, Berlin},\ \bibinfo {year} {2009})\BibitemShut
  {NoStop}%
\bibitem [{\citenamefont {Renlund}(2010)}]{Renlund:2010}%
  \BibitemOpen
  \bibfield  {author} {\bibinfo {author} {\bibfnamefont {H.}~\bibnamefont
  {Renlund}},\ }\href@noop {} {\enquote {\bibinfo {title} {Generalized
  p\'{o}lya urns via stochastic approximation},}\ } (\bibinfo {year} {2010}),\
  \Eprint {http://arxiv.org/abs/arXiv:1002.3716 [math.PR]} {arXiv:1002.3716
  [math.PR]} \BibitemShut {NoStop}%
\bibitem [{\citenamefont {Vatiwutipong}\ and\ \citenamefont
  {Phewchean}(2019)}]{Vatiutipong:2019}%
  \BibitemOpen
  \bibfield  {author} {\bibinfo {author} {\bibfnamefont {P.}~\bibnamefont
  {Vatiwutipong}}\ and\ \bibinfo {author} {\bibfnamefont {N.}~\bibnamefont
  {Phewchean}},\ }\href@noop {} {\bibfield  {journal} {\bibinfo  {journal}
  {Adv.Differ.Equ.}\ ,\ \bibinfo {pages} {276}} (\bibinfo {year}
  {2019})}\BibitemShut {NoStop}%
\end{thebibliography}%
\appendix

\section{SDE analysis of P\'{o}lya urn }
\label{sec:appendix}

We study the P\'{o}lya urn using the SDE for the reader's convenience.
We denote $D_{d}(t),S_{d}(t),Z_{d}(t)$ for the infinite memory kernel 
as $D_{\infty}(t)=t$,$S_{\infty}(t)=\sum_{s=1}^{t}X(s)$,$Z_{\infty}(t)=\sum_{s=1}^{t}X(s)/t$, respectively. To estimate the
expected value and the variance of $Z_{\infty}(t+1)$ with the condition
$Z_{\infty}(t)=z$, we decompose $Z_{\infty}(t+1)$ as
\[
Z_{\infty}(t+1)=Z_{\infty}(t)+\frac{1}{t+1}(X(t)-Z_{\infty}(t)).
\]
We estimate the conditional expected value and conditional variance
of $Z_{\infty}(t+1)$ as follows:
\begin{eqnarray}
E(Z_{\infty}(t+1)|Z_{\infty}(t)=z)-z&=&
\frac{\alpha-\theta z}{(D_{\infty}(t)+\theta)D_{\infty}(t+1)},
\nonumber \\
V(Z_{\infty}(t+1)|Z_{\infty}(t)=z)&=&
\frac{1}{D_{\infty}(t+1)^2}\cdot
\frac{\alpha+tz}{\theta+t}\cdot \frac{\beta+t(1-z)}{\theta+t}\nonumber.
\end{eqnarray}
We read the drift $A_{\infty}(t)$ and variance $B_{\infty}(t)$ from
them and  derive the SDE \cite{Gardiner:2009} as
\[
dZ_{\infty}=A_{\infty}(\alpha-\theta Z_{\infty})dt+B_{\infty}dW_t.
\]
$A_{\infty}(t),B_{\infty}(t)$ are defined as follows:
\[
A_{\infty}(t)=\frac{1}{D_{\infty}(t+1)(\theta+D_{\infty}(t))},
B_{\infty}(t)=\frac{B}{D_{\infty}(t+1)},
B=\sqrt{\frac{(\alpha+tz)(\beta+t(1-z))}{(\theta+t)^2}}.
\]

As $E(Z_{\infty}(t)|X(1)=x)=(\alpha+x)/(\theta+1)$ for the 
P\'{o}lya urn, we
take the limit $t\to\infty$ by replacing $z$ with $(\alpha+x)/(\theta+1)$
and approximate $B$ as
\[
B=\sqrt{\frac{(\alpha+x)(\beta+(1-x))}{(\theta+1)^2}}.
\]
The SDE is the Ornstein--Uhlenbeck
type SDE\cite{Gardiner:2009}.
The solution with the initial condition $Z_{\infty}(1)=X(1)=x$ is
\[
Z_{\infty}(t)=\frac{\alpha}{\theta}(1-e^{-\theta\int_{1}^{t}A_{\infty}(s)ds})
  +x\cdot e^{-\theta\int_{1}^{t}A_{\infty}(s)ds}
  +e^{-\theta\int_{1}^{t}A_{\infty}(s)ds}\int_{1}^{t}B_{\infty}(s)
  e^{\theta\int_{1}^{s}A_{\infty}(u)du}dW_s.
\]
We have $\int_{1}^{t}A_{\infty}(s)ds=
\frac{1}{\theta}\ln \frac{t(1+\theta)}{t+\theta}$,
and the solution is written as follows:
\[
Z_{\infty}(t)=\frac{\alpha}{\theta}\left(1-\frac{t+\theta}{t(1+\theta)}\right)
+x\cdot \frac{t+\theta}{t(1+\theta)}+\left(1+\frac{\theta}{t}\right)\int_{1}^{t}
\frac{B}{1+s}\cdot\frac{s}{s+\theta}dW_s.
\]
The conditional expected value of $Z_{\infty}(t)$ for $X(1)=x$ is
\[
E(Z_{\infty}(t)|X(1)=x)
=\frac{\alpha}{\theta}(1-\frac{t+\theta}{t(1+\theta)})
+x\cdot \frac{t+\theta}{t(1+\theta)}
\stackrel{t>>1}{=}
\frac{\alpha}{\theta}\cdot\frac{\theta}{1+\theta}+
x\cdot \frac{1}{1+\theta}.
\]
The ACF is then estimated as
\begin{eqnarray}
C(t)&=&E(X(t+1)|X(1)=1)-E(X(t+1)|X(1)=0) \nonumber \\
&=&E(P(X(t+1)=1|X(1)=1)-E(P(X(t+1)=1)|X(1)=0)  \nonumber \\ 
&=&\frac{D_{\infty}(t)(E(Z_{\infty}(t)|X(1)=1)-E(Z_{\infty}(t)|X(1)=0))}{D_{\infty}(t)+\theta}=\frac{1}{\theta+1}=C(1) \nonumber .
\end{eqnarray}
The result agrees with (\ref{eq:C_polya}).
We estimate the conditional variance of $Z_{\infty}(t)$ for $X(1)=x$ as
\[
V(Z_{\infty}(t)|X(1)=x)= B^2\left(1+\frac{\theta}{t}\right)^2
\cdot\frac{t-1}{(t+\theta)(1+\theta)}\simeq \frac{B^2}{1+\theta}
=\frac{1}{\theta+1}\cdot \frac{(\alpha+x)(\beta+(1-x)}{(\theta+1)^2}.
\]
Apart from the denominator of the prefactor,
it agrees with the exact result of (\ref{eq:V_polya}).

\end{document}